\documentclass[%
preprint,
 amsmath,amssymb,
 aps,
]{revtex4-1}
\usepackage{graphicx}
\usepackage{dcolumn}
\usepackage{bm}
\usepackage{amsmath}
\usepackage{mathtools}
\usepackage{subfig}

\begin{document}
\title{Choptuik's Critical Phenomenon in Einstein-Gauss-Bonnet Gravity}
\author{Sveta Golod and Tsvi Piran}
\altaffiliation{Racah Institute for Physics, The Hebrew University, Jerusalem, 91904 Israel}

\begin{abstract}
We investigate the effects of higher order curvature corrections to 
Einstein's Gravity on the critical phenomenon near the black hole threshold, namely the Choptuik phenomenon. 
We simulate numerically  a five dimensional spherically symmetric gravitational collapse of massless scalar field  in Einstein-Gauss-Bonnet (EGB) gravity towards a black hole formation threshold.
When the curvature is sufficiently large the   additional higher order terms,  affect the evolution of the whole system. Since high curvature characterizes the region when the critical behavior takes place 
this critical behavior  is  destroyed. Both the self similarity and the mass scaling relation  disappear.
Instead we find a different behavior near the black hole threshold, which depends on the coupling constant of the higher order terms. The new features include a change of the sign of the Ricci scalar on the origin which 
indicates changes in the local geometry of space-time, and never occurs in classical general relativity collapse, and oscillations with a constant rather than with a diminishing length scale. 
\end{abstract}


\maketitle

\section{Introduction}
Gravity is described by
Einstein's theory of general relativity (GR), which predicts the existence of 
Black holes  - trapped regions from which nothing can escape.
Black holes can form from regular initial data that do not contain a black hole already.
Isolated system in GR can end up in two qualitatively different states. 
Data that forms a black hole in the evolution and data that doesn't and in which the mass-energy disperses to infinity. 
A simplest possible model for such a system is the collapse of a spherically symmetric minimally coupled 
massless scalar field. 
The final fate - whether it collapses or not - depends on the ``strength'' of the initial data.


In a pioneering work Choptuik \citep{Choptuik} explored the transition between the two regimes. 
He discovered that the black hole threshold shows both surprising structure  and surprising simplicity. Universality, power-law scaling of the black hole mass, and scale echoing  have given rise  to the term ``critical phenomena''.\citep{Gundlach:gr-qc0210101,SorkinOren}.
Choptuik \citep{Choptuik} considered a one  parameter families of initial data describing a collapsing scalar field  (see figure \ref{Penrose} for the structure of space time)/ He have shown that 
for each family of initial data parametrized by $p$ (for example amplitude of the initial pulse) there exist a critical value $p_*$. 
For $p>p_*$ we have a supercritical collapse and a black hole forms. For $p<p_*$  the collapse is subcritical and the field disperses to infinity leaving a flat space. Choptuik gave a highly convincing numerical evidence that by fine-tuning the parameter $p$ to the threshold value $p_*$ an infinitely small black hole can be created.

\begin{figure}[ht!]
 \centering
\subfloat[\footnotesize Contours of the scalar field function $s$ 
vs $u$ and $v$ ]
{\label{fig:s_surf_0216}\includegraphics[width=0.5\textwidth]{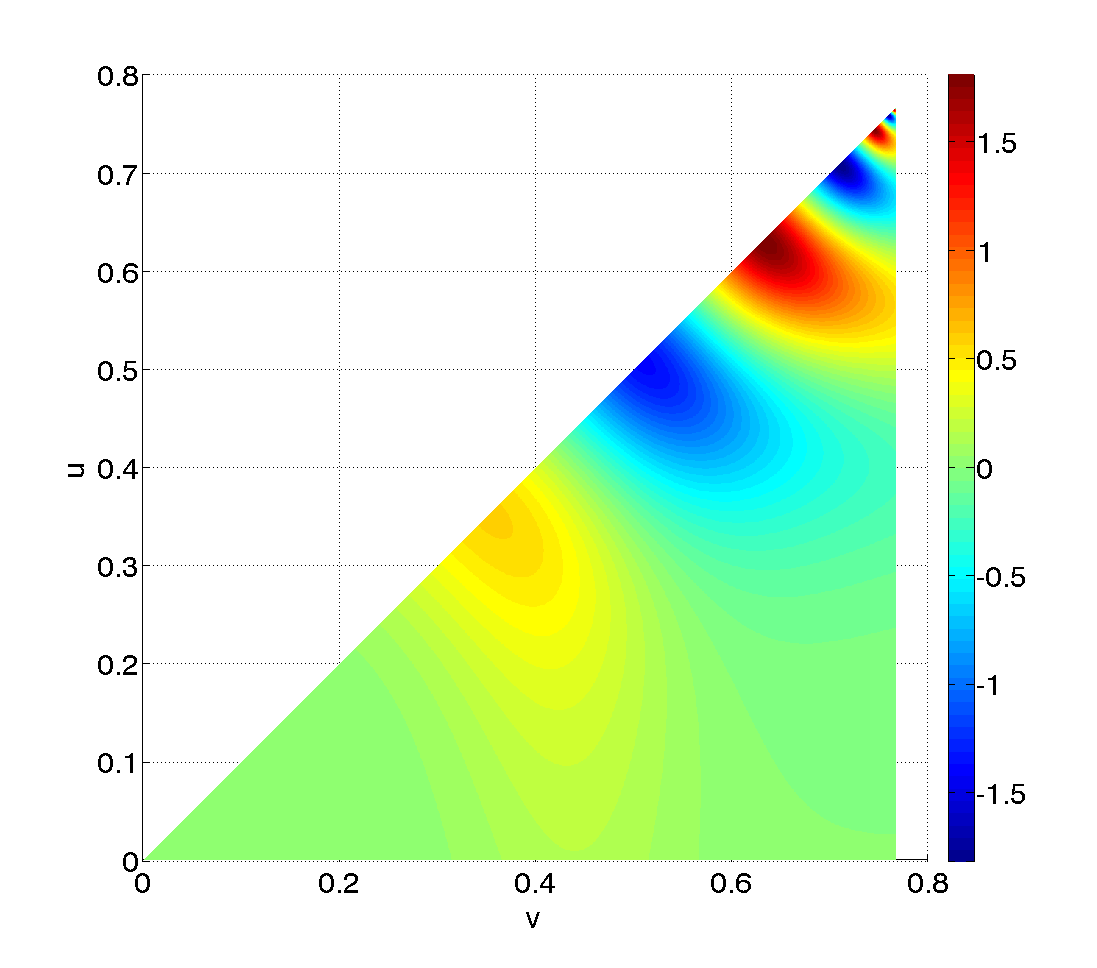}}
\subfloat[\footnotesize The field function $s$ at the origin  ($r=0$) vs $u$.]
{\label{fig:s_axis_0216}\includegraphics[width=0.5\textwidth]{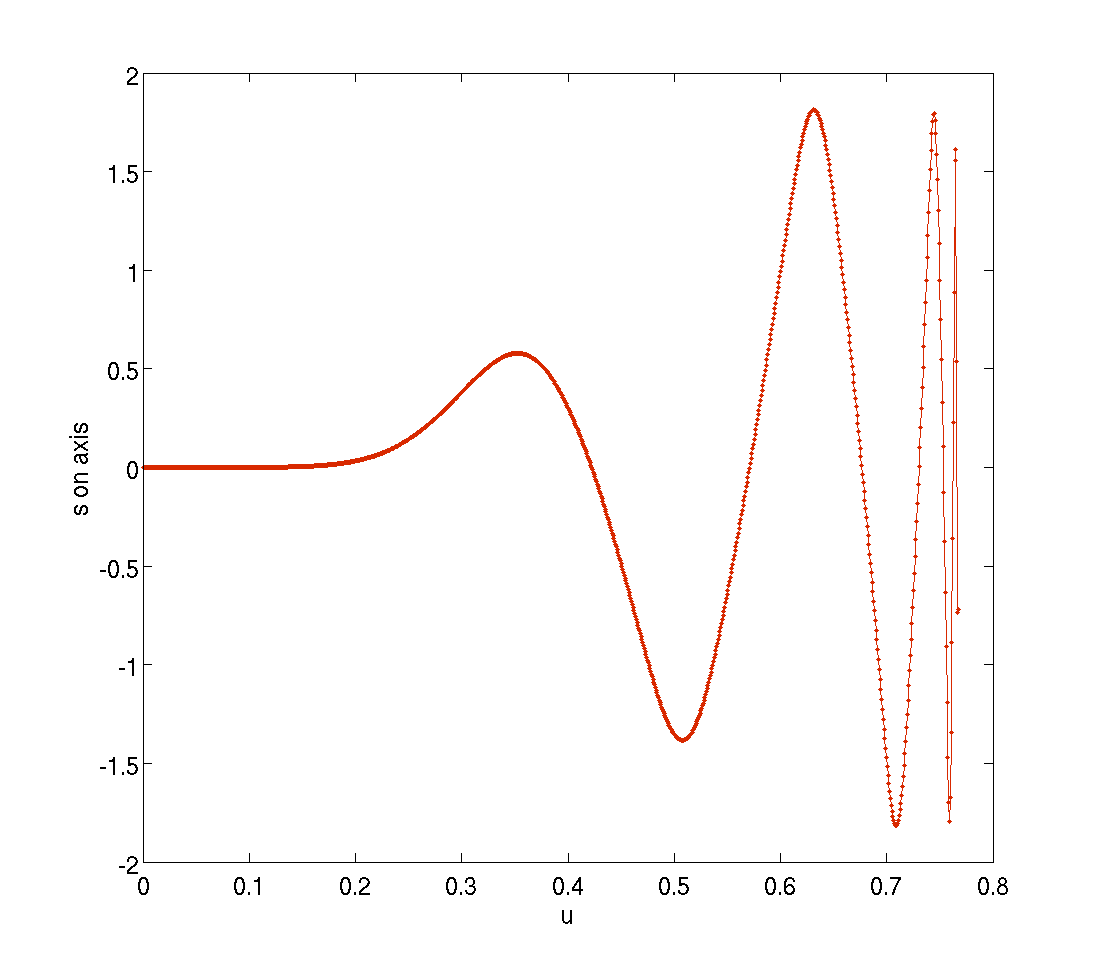}}
\caption{\footnotesize  The field function $s$ in classical GR in 5 dimensions, for a slightly subcritical collapse. 
The field oscillates with DSS pattern. The field pulsations decreasing in temporal and spatial scales. }
\label{fig:s_surf_0}
\end{figure}

The critical solution itself is universal. For a finite time in a finite region of space the spacetime 
converges to one and the same solution independent of the initial data. 
The critical solution is discretely self similar (DSS), namely, 
it is invariant under rescaling by a particular finite factor, or its integer powers.
Let $Z_*(r,t)$  be the critical solution (collectively for all the parameters - the scalar field and the metric), the critical 
solution is the same when rescaling space and time by factor $e^\Delta$:
\begin{equation}
Z_*(r,t)=Z_*(re^\Delta , te^\Delta ).
\end{equation}
The field and metric functions pulsate periodically with ever decreasing temporal and spatial scales, 
until a black hole forms in supercritical collapse, or the field disperses in subcritical collapse (see figure  \ref{fig:s_surf_0}). 
This universal phase ends when the 
evolution diverges towards a black hole formation or towards dispersion, depending on whether $p>p_*$ or not.
In supercritical collapse above  $p_*$ arbitrary small black holes are formed as $p \rightarrow p_*$ and the black hole mass scales as a power law: $ M \propto (p - p_*)^\gamma$,
where $\gamma$ is universal. It depends on the type of collapsing matter (and on the dimension) but it is independent of the initial data family.
Similar critical phenomena were found in many other types of matter coupled to gravity, with spherical symmetry and beyond it 
(See e.g. \citep{Gundlach:gr-qc0210101} for a review). 
The echoing period $\Delta$ and critical exponent $\gamma$ depend on the type of matter and on the dimension, but the phenomena appears to be generic.


\begin{figure}[ht!]
 \centering
\includegraphics[width=0.7\textwidth]{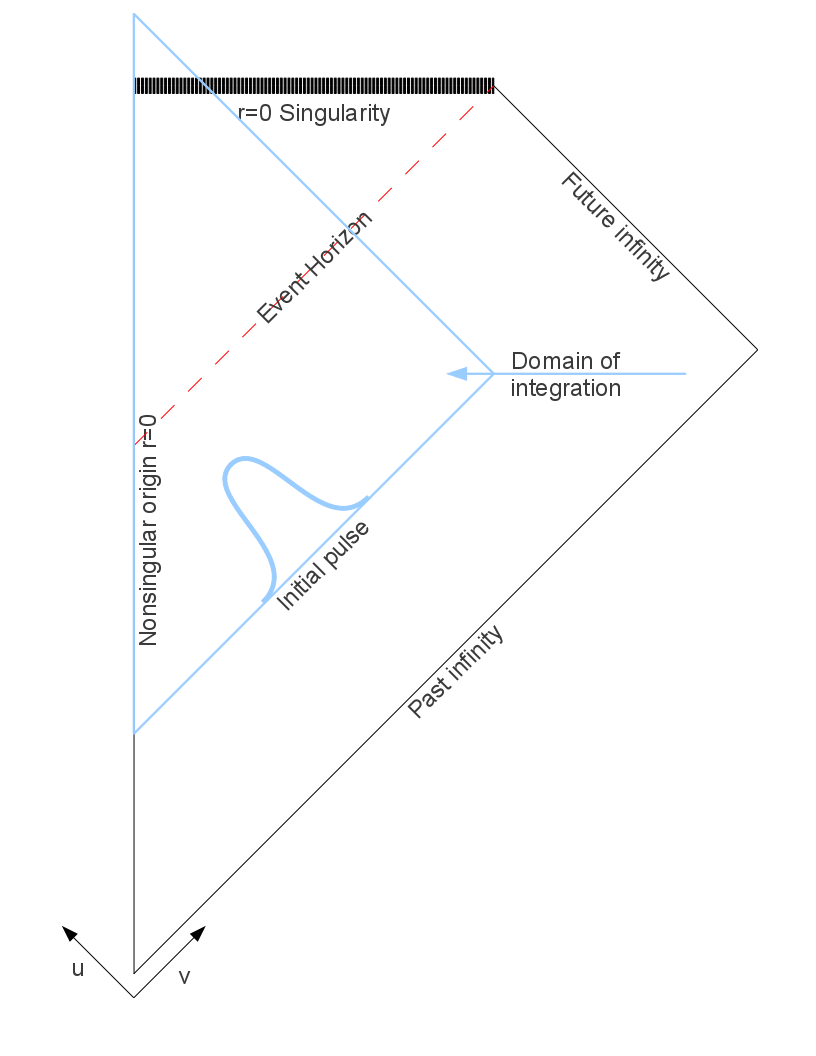}
\caption{\footnotesize 
The Penrose diagram in GR of the space time that is expected 
to form in a gravitational collapse of a shell of in-falling scalar field to a black hole.
When the shell is far away from the origin
the self gravitational effects are small. 
When it comes closer to the origin gravitational field becomes stronger.
If the field doesn't collapses to a black hole the diagram remains Minkowsky-flat, 
and the event horizon or the singularity don't exist, of course.
Light blue lines indicate the numerical domain of integration used in the current work. The initial hypersurface is a null 
ray. Since the field is massless it behaves light-like and it moves along null rays.}
\label{fig:penrose}
\end{figure}

Einstein's equations are derived from the Hilbert action, which is linear in the Ricci scalar - $R$. It is natural to expect that higher terms
in $R$ will appear in a more general theory and it is interesting to explore their possible role. To do so we have to explore a high curvature regions of space time where such terms are significant. 
Black hole formation is a natural place to do so as spacetime becomes highly curved as the matter fields  collapse. 
This behavior usually takes place near the singularity, which is typically hidden inside the black hole. However
they also appear near the  threshold for black hole formation when the Choptuik phenomenon take place. 
Therefore we explore here the gravitational collapse of a spherically symmetric massless scalar field with higher order corrections
to the Hilbert action.

Addition of even the simplest  $R^2$  term induces $4^{\rm th}$ order derivatives of the metric in the resulting
equations of motion.
The  original Einstein equations are second order in the metric and the additional generic $4^{\rm th}$ order corrections will govern 
the equations and change completely their character. To overcome this we consider a special case. 
Lanczos \citep{Lanczos, deruelle-2003} found a generalization of Hilbert's Lagrangian which is quadratic in the Riemann tensor and its contractions,  but its variation yields a system of equations that, like Einstein's, is second order in the metric derivatives. This correction to the Lagrangian, called the  Lanczos Lagrangian or the Gauss-Bonnet term \citep{kobayashiNomizu,PatersonLondon}, is given by:
\begin{equation}\label{eq:L_GB}
 L_{GB} \equiv  R^2-4R_{\mu\nu}R^{\mu\nu}+R_{\mu\nu\rho\sigma}R^{\mu\nu\rho\sigma}.
\end{equation}
The combined theory that includes the Hilbert Lagrangian and the Gauss-Bonnet term is called Einstein-Gauss-Bonnet (EGB) gravity.
In four dimensions the Gauss-Bonnet term is a pure divergence, just like 
Hilbert's Lagrangian in two dimensions, and it does not contribute
to the field  equations in four  dimensions. To overcome this we consider here gravitational collapse in  5 dimensional space-time, which is the simplest system in which the Gauss-Bonnet term contributions can affect the evolution\footnote{Interestingly, the Gauss-Bonnet term in the Lagrangian,  defined in equation (\ref{eq:L_GB}),
is the higher curvature correction to general relativity that naturally arises as the next leading order of the 
$\alpha$' expansion of heterotic superstring theory, where $\alpha$' is the inverse string tension.
\citep{GrossSloan,BentoBertolami,Zwiebach1985315,MetsaevTseytlin}.} .


We present here the results of a numerical investigation of  the influence of higher order curvature correction, namely the Gauss-Bonnet term, on the 
properties of spherically symmetric scalar field collapse in 5 dimensions. In particular we focus  on the behavior of the
Choptuik critical phenomenon. 
The structure of the paper is as follows:
in \S \ref{model} we describe the overall model and the basic equations. In \S \ref{sec:Numerical_Method} we discuss the numerical structure and the numerical difficulties that arise in the calculations. Simulation results and their discussion are presented in \S \ref{sec:RESULTS}.

 \section{The Model and the Basic Equations}
\label{model}
We consider the collapse of a spherically symmetric massless scalar field, $\phi$ in 5 dimensional space time described by the metric  $g_{\alpha\beta}$. 
We use units in which $G=c=1$. 

\subsection{The Metric}\label{sec:theMetric}
The scalar field is massless and it propagates along the light cone. 
Hence we describe the 5 dimensional asymptotically flat spacetime in double null coordinates:
\begin{equation}\label{null_metric}
 ds^2=-a^2(u,v)dudv+r^2(u,v)d\Omega^2_{3},
\end{equation}
where our coordinates are: $(u,v,\varphi,\theta,\gamma)$  and
$d\Omega^2_{3}=\left[ sin^2\gamma \left(d\theta^2 + sin^2\theta d\varphi^2 \right) + d\gamma^2 \right] $ is the metric on 3 dimensional unit sphere (see \citep{Kunstatter+12,Taves+12} for an alternative Hamiltonain formulation of this problem).
The coordinate 
$u$ is the retarded time coordinate and a constant $u$ describes an outgoing null trajectory. 
Similarly $v$ is the advanced time coordinate and surfaces with a constant $v$ are the ingoing null trajectories.
$r\equiv r(u,v)$ is the area coordinate and $r=0$ is the origin of the spherical symmetry. 
This  definition of the metric is unique only up to a change of variables of the form 
$v\rightarrow \tilde{v}(v)$, $u\rightarrow \tilde{u}(u)$. This gauge freedom will be fixed later by the choice of the initial conditions.

\subsection{The Scalar Field}
The Lagrangian density of the field is:
\begin{equation}
 L=-\frac{1}{2}\phi_{;\alpha}\phi^{;\alpha}, 
\end{equation}
and the corresponding equation of motion  is:
\begin{equation}\label{field_eq}
 \phi^{;\sigma}_{\ \ ;\sigma}=0.
\end{equation}
The energy-momentum tensor is given by:
\begin{equation}\label{energy-momentum_tensor}
 T_{\alpha\beta}=\phi_{,\alpha}\phi_{,\beta}-\frac{1}{2}g_{\alpha\beta}\phi_{,\sigma}\phi^{,\sigma}
=
  \begin{pmatrix}
    T_{uu}	&T_{uv}	&0			&0			&0 \\
    T_{vu}	&T_{vv}	&0			&0			&0 \\
    0		&0	&T_{\varphi\varphi}	&0			&0 \\
    0		&0	&0			&T_{\theta\theta}	&0 \\
    0		&0	&0			&0			&T_{\gamma\gamma}
  \end{pmatrix} 
=
 \begin{pmatrix}
    \phi^2_{,u}	&0		& 				& 				&  \\
     0		&\phi^2_{,v}	& 				& 				&  \\
     		& 		&sin^2\theta T_{\theta\theta}	& 				&  \\
     		& 		& 				&sin^2\gamma T_{\gamma\gamma}	&  \\
     		& 		& 				& 				&\frac{2r^2}{a^2}\phi_{,u}\phi_{,v}
  \end{pmatrix} . 
\end{equation}

\subsection{The equations of Motion}\label{section_eq_motion}
The overall action that includes the Hilbert action and the Gauss-Bonnet term is:
\begin{equation}
S=\int d^nx\sqrt{-g}\left[ \frac{1}{16 \pi}\left(R+\alpha L_{GB}
 \right) 
\right] +S_{matter},
\end{equation}
where 
$\alpha$ is the coupling constant and it has dimensions of $(length)^2$. 
The value of the coupling constant $\alpha$ is unknown, 
we assume that $0<\alpha <<1$.
As we are interested in $\alpha$ values which are significant and influence the solution. Therefore we will look for values that are large enough so that  the correction term which is of order $\alpha R^2$ is comparable or larger then  the Hilbert term $R$.  Namely we will be interested in cases where  $\alpha R > 1$. 

The  corresponding gravitational field equations  are:
\begin{equation}\label{new_Einstein_eq}
 G_{\mu\nu}+\alpha H_{\mu\nu}=\kappa_n^2 T_{\mu\nu},
\end{equation}
where $ G_{\mu\nu}$ is Einstein tensor:
\begin{equation}
 G_{\mu\nu}=R_{\mu\nu}-\frac{1}{2}Rg_{\mu\nu},
\end{equation}
$T_{\mu\nu}$ is the energy-momentum tensor  given in (\ref{energy-momentum_tensor}), and
\begin{equation}
H_{\mu\nu}\equiv 2\left[ R R_{\mu\nu}-2R_{\mu\alpha}R^{\alpha}_{\ \nu}-2R^{\alpha\beta}R_{\mu\alpha\nu\beta}
+R_{\mu}^{\ \alpha\beta\gamma}R_{\nu\alpha\beta\gamma}
\right] 
-\frac{1}{2}g_{\mu\nu}L_{GB}.
\end{equation}
We convert the equations to a set of first order differential equations. To do so we define:
\begin{align}
&s\equiv  \sqrt{4\pi G}\phi \\ 
&z\equiv  s_{v} \label{z}\\ 
&w\equiv  s_{u} \label{w}\\  
&d\equiv \frac{a_{v}}{a} \label{d}\\
&f\equiv  r_{u} \label{f}\\
&g\equiv  r_{v} \label{g},
\end{align}
where  $Z_{\mu}\equiv  \frac{\partial Z}{\partial x^{\mu}}$.
We obtain, using  equation (\ref{new_Einstein_eq}) four
independent first order  equations:
\begin{equation}\label{f_u}
f_u=2f\frac{a_u}{a}-\frac{2 G r a^2 w^2}{3\left[a^2+4\alpha \eta \right] } \quad ;~   \underline{uu} {\rm ~ component};
\end{equation}
\begin{equation}\label{g_v}
g_v=2gd-\frac{2 G r a^2 z^2}{3\left[a^2+4\alpha \eta \right] } \quad ;~ \underline{vv} {\rm ~ component};
\end{equation}
\begin{equation}\label{f_V}
f_v=g_u=-\frac{a^2 r \eta}{2\left[a^2+4\alpha \eta \right] } \quad ;~  \underline{uv} {\rm ~ component};
\end{equation}
and 
\begin{align}\label{d_u}
d_u =& \frac{-9a^2 \eta[(4\alpha\eta)^2-3 a^4]+4 a^2 G zw[-9(a^2+4\alpha \eta)^2 +32 a^2 G zw\alpha]} 
{36(a^2+4\alpha \eta )^3} \quad  ;~ \frac{a^2}{2r^2}\underline{\gamma\gamma}+\underline{uv} {\rm ~ components}.
\end{align}
The scalar field equation derived from (\ref{field_eq}) is:
\begin{equation}\label{w_v}
 w_v=z_u=-\frac{3}{2r}(gw+fz).
\end{equation}
We have defined here an auxiliary function 
$\eta$:
\begin{equation}\label{eta}
\eta\equiv \frac{a^2+4fg}{r^2}.
\end{equation}
This is useful to stabilize the numerical solution, as 
explained later in Section \ref{sec:Numerical_Method}.
The evolution equation for $\eta$ is:
\begin{equation}\label{eta_v}
 \eta_v=-\frac{2\eta(g-rd)}{r}-\frac{2a^2 (3g\eta+4fz^2)}{3r(a^2+4\alpha\eta)}.
\end{equation}

Equations - (\ref{f_u}), (\ref{g_v}), (\ref{f_V}), (\ref{d_u}),
(\ref{w_v}), (\ref{eta_v})  - together with
(\ref{z}), (\ref{w}), (\ref{d}), (\ref{f}) and (\ref{g}) are   a set of eleven coupled first 
order differential equations that describe the system. 
Since there are only nine functions - $r,a,s,f,g,z,w,d$ and $\eta$ - two equations are redundant and are simply the
usual constraint equations. 
We choose those to be Eqs.  (\ref{f}) and (\ref{f_u}). They are not evolved in the integration, 
but they are monitored to verify that they are indeed satisfied during the evolution.

The original GR  equations are invariant under rescaling.
However, the coupling constant $\alpha$ has units of length squared $[L^2]$.  It's introduction 
to the system destroys this scale invariance. 
Under the rescaling: $r\longrightarrow \rho r$ the equations behave as:
\begin{equation}\label{eq:f_u_rescaling}
f_u=2f\frac{a_u}{a}-\frac{2 G r a^2 w^2}{3\left[a^2+4\alpha \eta \right] } \quad 
\longrightarrow \quad
\frac{1}{\rho}f_u=\frac{1}{\rho}2f\frac{a_u}{a}-\frac{1}{\rho} \frac{2 G r a^2 w^2}{3\left[a^2+\frac{4\alpha \eta}{\rho^2} \right] }
\end{equation}
\begin{equation}
g_v= 2gd- \frac{2 G r a^2 z^2}{3\left[a^2+4\alpha \eta  \right] } \quad
\longrightarrow \quad
\frac{1}{\rho}g_v=\frac{1}{\rho}2gd-\frac{1}{\rho}\frac{2 G r a^2 z^2}{3\left[a^2+\frac{4\alpha \eta}{\rho^2} \right] },
\end{equation}
\begin{equation}
f_v=- \frac{a^2 r \eta}{2\left[a^2+{4\alpha \eta} \right] } \quad 
\longrightarrow \quad
\frac{1}{\rho}f_v=-\frac{1}{\rho} \frac{a^2 r \eta}{2\left[a^2+\frac{4\alpha \eta}{\rho^2} \right] },
\end{equation}
\begin{eqnarray}
d_u = \frac{-9a^2 \eta[(4\alpha\eta )^2-3 a^4]+4 a^2 G zw[-9(a^2+4\alpha \eta )^2 +32 a^2 G zw\alpha  ]} 
{36(a^2+{4\alpha \eta} )^3}
 \quad 
\longrightarrow  \nonumber \\
\quad
\frac{1}{\rho^2}d_u =\frac{1}{\rho^2} \frac{-9a^2 \eta[(\frac{4\alpha\eta}{\rho^2})^2-3 a^4]+4 a^2 G zw[-9(a^2+\frac{4\alpha\eta}{\rho^2})^2 + \frac{32 a^2 G zw\alpha}{\rho^2} ]} 
{36(a^2+\frac{4\alpha \eta}{\rho^2} )^3}.
\end{eqnarray}

Obviously the scale invariance is broken. At smaller scales  ($\rho \rightarrow 0$) the non-invariant elements, which rescale as $\rho^{-2}$, become dominant and govern the equations. This  leads to  the deviations from the classical GRB behaviour that we demonstrate numerically later. In particular the deviations from the GR behavior happen near the critical point where the curvature is large and the non scale invariant Gauss-Bonnet terms are large. These terms eventually destroy the classical Choptuik phenomenon both when black holes form and when they don't. 
This happens independently of the value of $\alpha$.  

\subsection{The Boundary Conditions}\label{sec:boundaryConditions}

Regularity and differentiability at the origin $r=0$   require the following
boundary conditions:
\begin{equation}
\begin{array}{lll}
 g=-f=\frac{a}{2}, & \partial_r s=0, & \partial_r a=0.
\end{array}
\end{equation}
These conditions imply  
\begin{equation}
a_v=a_u \quad ; \quad w=z \quad ; \quad \eta=\dfrac{a}{12\alpha}(-3a+\sqrt{9a^2+48\alpha z^2})
\end{equation}
at the origin. 

{A second set of boundary conditions is set implicitly at infinity.  This is trivial in GR where a Schwartzchild
has naturally an asymptotically flat space time. However, \cite{BoulwareDeser85} have shown that a Schwartzchild black hole in EGB  gravity can have either asymptotically flat or AdS structure. Since we consider only a finite region of space time (see figure \ref{fig:penrose}) we don't examine here to which to the two branches the collapsing black hole will lead. }

\subsection{The Ricci Scalar}
The Ricci scalar curvature, $R$, is given by:
 \begin{equation}\label{R}
 R=8\frac{-9 a^6 w z + 432 \eta^4 \alpha^3 +72a^2 \eta^2 (\eta-2wz)\alpha^2 +a^4(32w^2 z^2-72w z \eta -27\eta^2)\alpha}
{9 a^2 (a^2+4\alpha \eta)^3}.
\end{equation} 
The Ricci curvature describes the local geometry of the space-time. The value of the Ricci scalar 
at  the origin is of special interest:
 \begin{equation}\label{R_axes}
 R(r=0)=\frac{16z^2}{a \sqrt{9a^2+48z^2\alpha}}+\frac{5}{\alpha}\left(\frac{a}{\sqrt{a^2+\frac{16}{3}z^2\alpha}}-1 \right). 
\end{equation}
If $\alpha=0$ the Ricci scalar at the center is always negative: $-\dfrac{8z^2}{a^2}$. For $\alpha\neq0$ the situation
is more complicated. The Ricci scalar is negative while the following condition is satisfied:
\begin{equation}\label{eq:ricciCondition}
\alpha<\dfrac{45a^2}{16z^2}.
\end{equation}
The simulations show that along the evolution of the collapsing field system, 
the metric function - $a(u,v)$ approaches zero and $z(u,v)$,  the field derivative,
is growing to a very big values, the closer we are to the critical amplitude, the larger is the value 
that $z$ approaches. Therefore, the condition of eq.(\ref{eq:ricciCondition}) is violated at some point
and the Ricci scalar changes sign. This change of sign heralds the deviation from the classical behavior.

\subsection{The Black Hole Mass}\label{sec:black hole_mass}
The analysis of mass scaling relation in the critical phenomenon requires a function for a black hole mass. 
The ADM mass of a black hole in higher dimensional GR \citep{Tangherlini} is:
\begin{equation}
 M=\frac{(D-2)A_{D-2}}{16\pi G_D}r_s^{D-3},
\end{equation}
when $r_s$ is Schwarzschild radius, $D$ - the dimension, $G_D$ is the D-dimensional Newton constant
and $A_{D-2}$ is the area of a unit sphere: $A_{D-2}=\dfrac{2 \pi^{\frac{D-1}{2}}}{\Gamma(\frac{D-1}{2})}$.
{However, in EGB there is an additional term and for  $D=5$ dimensions the ADM mass is
\citep{BoulwareDeser85,Wheeler86,Cai02,ToriiMaeda05} :
\begin{equation}\label{ADM_mass}
 M=\frac{3\pi}{8G}r_s^{2}(1 + \frac{2 \alpha}{r_s^2}) .
\end{equation}
The second term in this equation implies that as the size of the black hole decreases
($r_s \rightarrow 0$) its mass in EGB approaches a constant
positive value,   $M\rightarrow M_0 = 3 \pi \alpha /4 G $ \citep{Cai02,ToriiMaeda05}. This implies that there is a
mass gap and all black holes (for $\alpha >0$) must have an ADM  mass larger than  $M_0$. One can resort to a different definition of the black hole's mass and instead of using the ADM mass one can calculate the mass of the apparent (trapping) horizon using an EGB quasi-local mass \citep{Maeda06,MaedaNozawa08}.  Avoiding this problem we will consider, for simplicity,  in the following the scaling of the black hole's radius instead of the scaling of the black hole's mass.

}

We define the critical exponent $\gamma$ such that $|p-p_*|^\gamma$ has a dimension of length.
Instead of examining the dependence of the ADM mass on $p$ we will examine the dependence of 
the black hole's radius, $r_s$. We expect following the GR case to find $r_s\propto(p-p_*)^{\gamma}$.


\subsection{Initial  conditions}\label{sec:initialAndBoundaryConditions}
We turn to discuss the initial  conditions. 
We consider here the gravitational collapse of a shell of in-falling scalar field.
Figure \ref{fig:penrose} shows the domain of the current 
numerical work embedded inside an expected Penrose diagram. 
The  u-v plane is covered by a two dimensional grid, as described in Figure \ref{integration_domain}.
The origin $r=0$ is included in the domain, and it is chosen to be at $u=v$.
Therefore the relevant part of u-v space is  $v>u$.
Our metric is defined up to a coordinate
transformation, as it was mentioned in section \ref{sec:theMetric}. 
This gauge freedom is fixed by specifying the metric functions on the initial hypersurface, 
an initial ray with a constant retarded time $u=u_i=0$.
In flat - Minkowsky - space-time the conventional definition of the null 
coordinates is $u=t-r$ and $v=t+r$. Since we are dealing with spherical shell,
the space-time is flat in two regions - inside the shell and at
asymptotically large radii. The integration starts far away from the event horizon and therefore  the metric is nearly flat.
Thus we define the area coordinate $r$ along the initial null 
surface $u=u_i=0$ as in a flat space time $r\equiv \dfrac{v}{2}$. The fact that the spacetime is only approximately Miknowski 
is  pronounced by the deviation of other  metric function - $a(u,v)$ - from its flat-space value  ($a=1$).
In addition, we set $a=1$ at the origin, $r=0$, at one point $u=v=0$. 
From here we can obtain all the other functions on the initial hypersurface by integration from the origin.

The exact shape of the initial scalar field is unimportant as the Choptuik behavior is universal and independent of this shape. We choose 
the initial scalar field profile along an outgoing hypersurface $u=u_i=0$  to be a Gaussian:
\begin{equation}
 s(u=0,v)=p \textbf{ } exp\left[-\left(\frac{v-v_c}{\sigma}\right)^2\right],
\end{equation}
where the constants $v_c$ and $\sigma$ determine the initial position and width of the shell and the constant $p$ is the amplitude of the pulse.
$p$ is the strength parameter of the initial data and it is the dynamical parameter that we vary to explore the 
 Choptuik phenomenon.

Having specified the  functions $r$ and $s$ on the initial hypersurface,
we can derive analytically $z$ and $g$, which are simply the derivatives
$z=s_v$ and $g=r_v=1/2$. All the other functions on the initial hypersurface 
 - $f$, $a$, $w$, $d$ and $\eta$ - are obtained by integrating the appropriate 
equations from the origin.

\section{Numerical Methods}\label{sec:Numerical_Method}
We follow here  the methods developed by Sorkin and Oren \citep{SorkinOren}. However, as discussed later, further steps, including the addition
of the variable $\eta$ are needed here to stabilize the code near the origin. The problem arises, because of the stronger non linear behavior of the Gauss-Bonnet terms. 

\subsection{The Integration Sceme}

\begin{figure}[!ht]
 \centering
\includegraphics[width=5.00in]{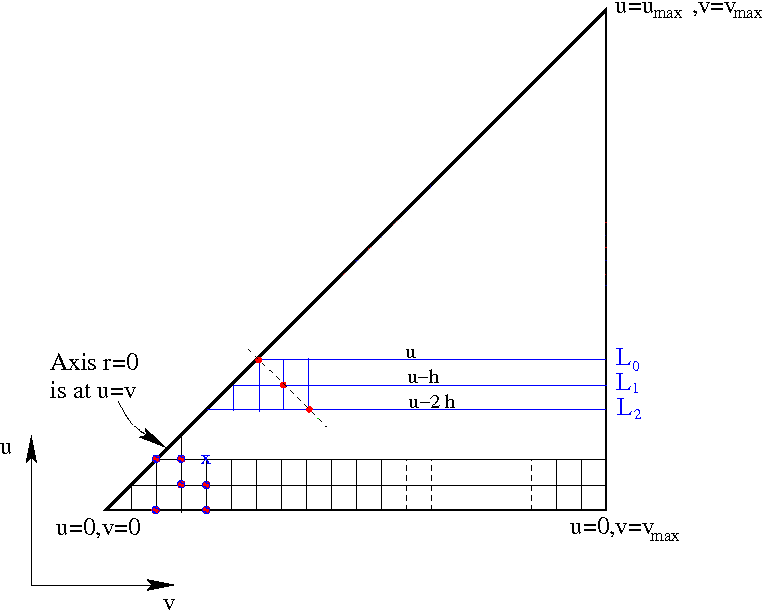}
\caption{\footnotesize The domain of integration. 
The calculations employs   the two previous lines $L_1$ and $L_2$ 
in addition to the line, $L_0$, that is currently being solved. The boundary conditions on $r=0$ involving $\partial_r$ 
are implemented using the 3-point derivatives along the diagonal line. Smoothing of some functions
near $r=0$ (at a point marked by cross) is done using past light cone points (marked by circles).
Figure taken from \citep{SorkinOren}.}
\label{integration_domain}
\end{figure}

Our domain of integration is a equilateral right angle triangle in a u-v plane: $0\le u \le u_{max}$,  $0\le v \le v_{max}$
and $v \ge u$.  $v_{max}=u_{max}$  is chosen to cover the interesting relevant region. 
An illustration of the domain is sketched in figures \ref{fig:penrose} and \ref{integration_domain}. 
The simplest computational cell is square with grid spacing $h_u=h_v=h$. Triangular cells 
near the origin ($u=v$)  are treated separately. 


The integration begins from the lower line of constant $u=0$ and propagates to the next line. 
Once the solution along an outgoing hypersurface with constant $u$ value, $u=U-h$, is known, 
$d$ and $z$ are propagated to the next line, $u=U$, using
equations (\ref{d_u}) and (\ref{w_v}) correspondingly. 
Then equations (\ref{w_v}), (\ref{f_V}), (\ref{g_v}), (\ref{z}), 
(\ref{g}), (\ref{d}) and (\ref{eta_v}) are integrated using a  fourth order Runge-Kutta algorithm from the origin outward
along $v$ to obtain the functions $w$ ,$f$, $g$, $s$, $r$, $a$ and $\eta$, 
respectively. The remaining equations are not used directly, 
but they must be satisfied and are used to test the numerical solution.

The first points near the origin are treated separately. The region near the origin is unstable. The instability arises near the origin where  discretization errors are amplified, especially in source terms that involve a division by $r$.
To resolve this problem we take several steps. First we introduce the new variable 
$\eta\equiv \dfrac{a^2+4fg}{r^2}$. This, unnecessary from the first glance, variable
is an algebraic combination of others.  It appears in every source function. It includes a division
by $r^2$ which is very sensitive to errors near the origin. 
The independent evolution of $\eta$  using equation (\ref{eta_v}) help stabilizing the source functions. 
In addition we have to take a few more steps:
We use at points near the origin a more stable, second order Runge-Kutta algorithm.
\begin{itemize}
\item Instead of integrating the function functions $f$ and $w$ along $v$ we evaluate these functions
using a Taylor expansion, e.g.:
\begin{equation}
f(v)=f(v_{0})+dv f_v(v_{0})+\frac{(dv)^2}{2}f_{vv}(v_{0})+O(dv)^3,
\end{equation}
when $v_0$ is the $v$ value on the origin and $dv\equiv v-v_0$.
\item Additionally we smooth  the functions $z$ and $d$. First we evaluate the function, at some point $P$, 
then its value is smoothed with the values of the same function on points on the past light cone of $P$ (see Figure \ref{integration_domain}). 
For example, for the function $z$ at the point marked by cross in Figure \ref{integration_domain} the new, smoothed, $z$ is calculated according to:
\begin{equation}
z_{new}=\left( \omega \cdotp z_{e} + \sum\limits_{i=i}^3 z_i \right) \frac{1}{3+\omega},
\end{equation}
where $z_{e}$ is the value obtained from the evolution equation, $z_i$ are the extrapolated values of $z$
along the 3 directions of the past light cone: $z_i(u)=2z(u-h)-z(u-2h)$,
and $\omega$ is a weight parameter, 
which is varying for different functions and different code parameters, but typically $\omega\propto0.1$.
\end{itemize}

Every one of those actions on its own stabilizes the integration but 
is insufficient to keep the code completely stable till the end.   For the classical GR evolution, with $\alpha=0$, $\eta$ is not needed,
however it is essential for the more general evolution. 
This combination  prevents the code from crashing at least for low enough values of $\alpha$. However, a strong penalty is paid as the combination and in particular the smoothing reduces the convergence rate of the code to a linear order.  Using this algorithms
we are able to get a stable and convergent evolution for small values of $\alpha$. 
However,  the code still becomes unstable for large values of  $\alpha$, usually when the  field amplitudes that are  close to the critical one. 

\subsection{Numerical tests}

We performed a series of simulations with  step sizes $h$, $\frac{h}{2}$ and $\frac{h}{4}$
in order to determine the accuracy of the numerical method.
If the numerical solution  converges, the relation between the different
numerical solutions and the real one will be:
\begin{equation}
 F_{real}=F^h+O(h^n),
\end{equation}
where $n$ is the order of convergence and $F^h$ is the numerical 
solution with step size $h$. 
For halved step sizes the error is reduced correspondingly:
$
  F_{real}=F^{h/2}+O((\frac{h}{2})^n)
$
and 
$
   F_{real}=F^{h/4}+O((\frac{h}{4})^n).
$
By defining (as in \citep{SorkinPiran2001}):
$
  c1\equiv F^h-F^{h/2}
$
and
$
  c2\equiv F^{h/2}-F^{h/4},
$
we find the convergence rate:
\begin{equation}\label{n}
n=log_2\left( \frac{c1}{c2}\right).
\end{equation} 
\begin{figure}[ht!]
 \centering
\subfloat[\scriptsize The convergence rate - n.]
{\label{n_convergence_rate}\includegraphics[width=0.48\textwidth]{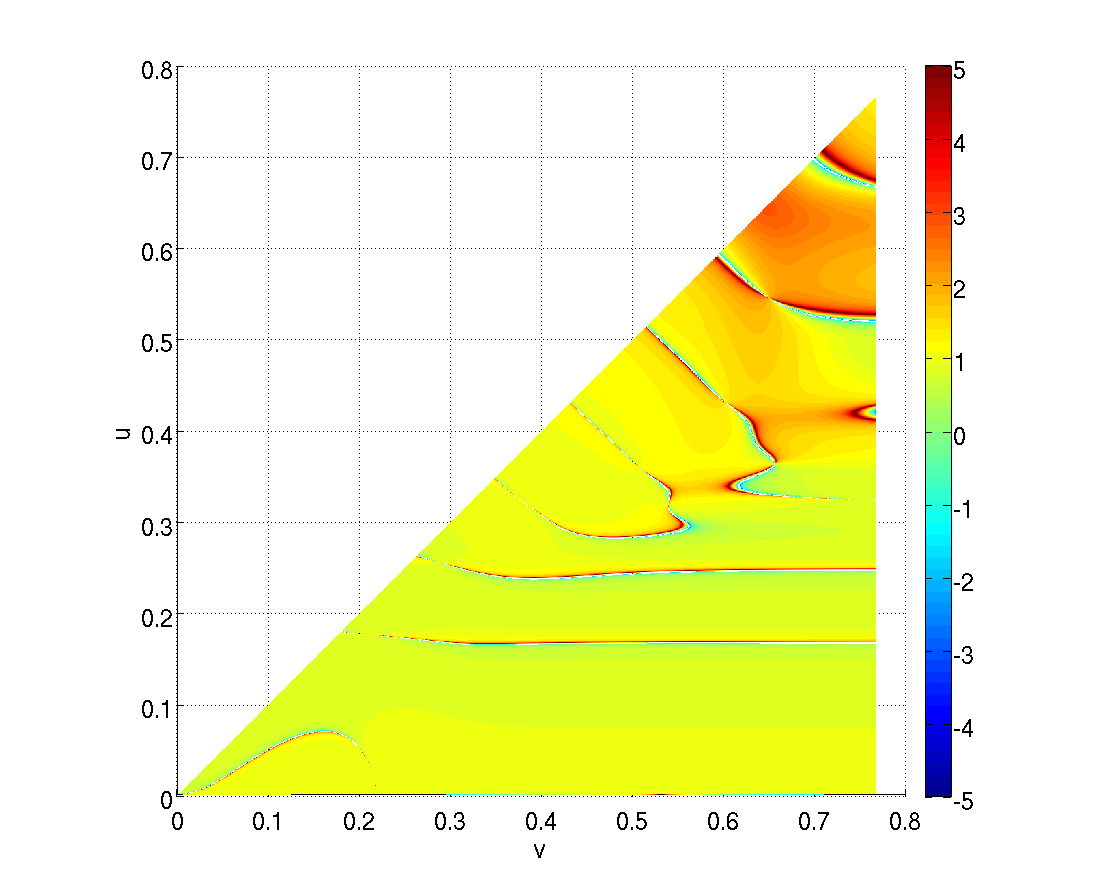}}
\subfloat[\scriptsize  One dimentional convergenge along a constant u-ray, $u=0.4$.
The convergence rate - n (upper panel) and the filed function - s for different grid densities
(lower panel)]
{\label{s_n_convergence}\includegraphics[width=0.48\textwidth]
{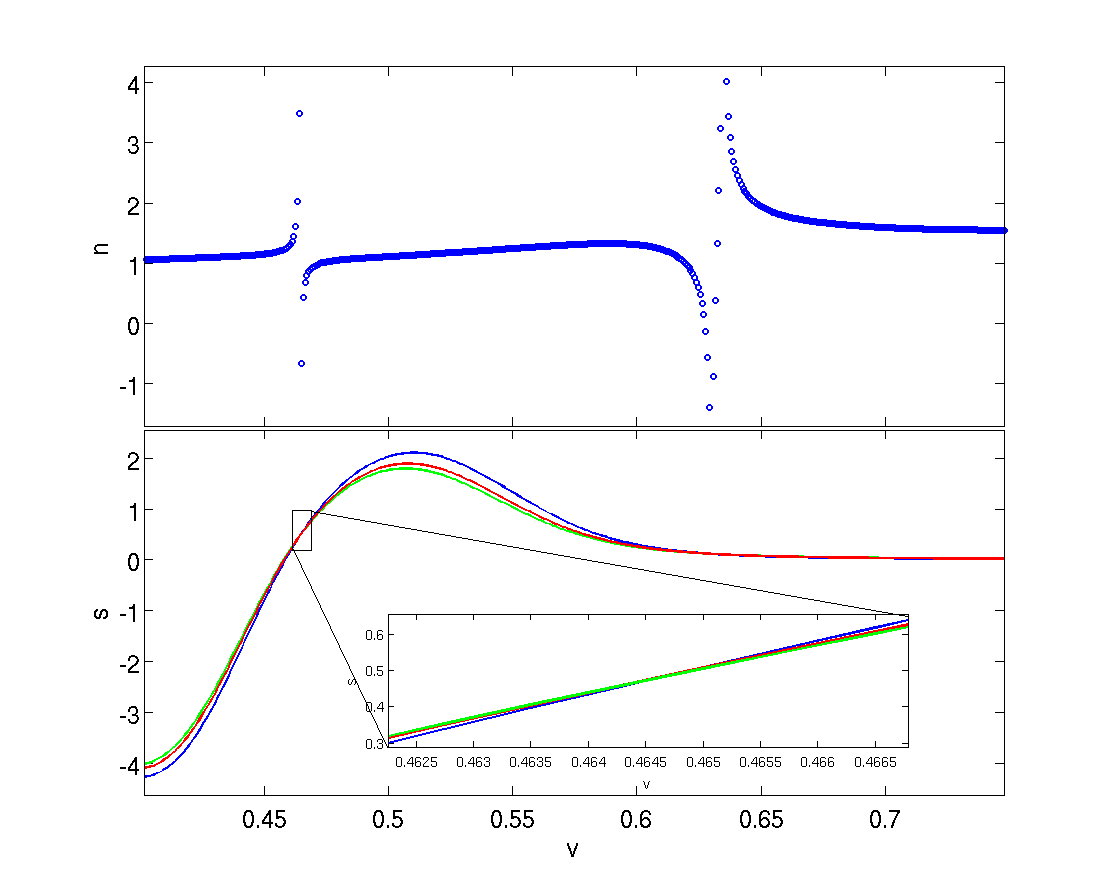}
}
\caption{\footnotesize 
Convergence for a slightly subcritical run with $\alpha=10^{-4}$.
The convergence rate - $n$ - in panel (a) is derived from the field function $s(u,v)$, with 
$2^{16}$, $2^{17}$ and $2^{18}$ grid points in $u$ and $v$ directions 
(blue, red and green lines correspondingly in $s$ plot on panel (b)).
In panel (b) the divergence of $n$ around $v\sim0.46$ and $v\sim0.63$ 
is caused by $s$ lines crossing, as shown in the zoom window.
}
\label{n_and_s_convergence}
\end{figure}

Figure \ref{n_convergence_rate} depicts $n$. 
The convergence rate is approximately linear $n\approx1$ or higher
for almost the whole domain. However, $n$ diverges at  some points.
This arises from crossing of $s=const.$ lines as can be shown in 
figure \ref{s_n_convergence}.
The upper panel of this figure shows a one dimensional projection of $n$ 
along constant u-ray: $u=0.4$, and
the lower panel shows the corresponding field function $F=s(u,v)$ 
along the same constant $u$ ray for different grid densities.

\section{Results}\label{sec:RESULTS}
\begin{figure}[ht!]
 \centering
\subfloat[\footnotesize Subcritical collapse, $p<p_*$.]
{\label{fig:noblack hole}\includegraphics[width=0.35\textwidth]{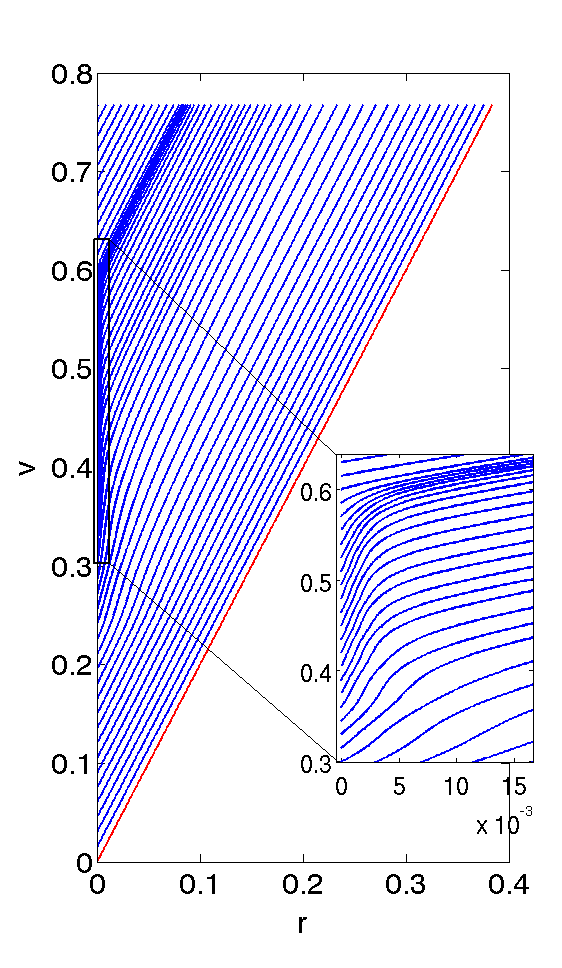}}
\subfloat[\footnotesize Supercritical collapse, $p>p_*$.]
{\label{fig:black hole}\includegraphics[width=0.35\textwidth]{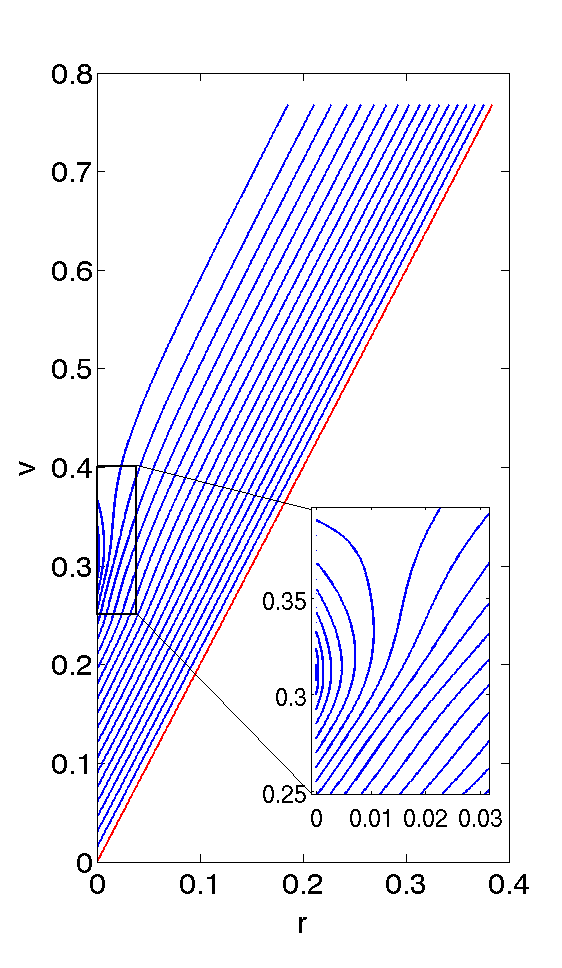}}
\caption{\footnotesize Outgoing null rays: $v$ vs $r$. 
Both plots have the same  parameters except for the initial field amplitude: 
$\alpha=10^{-5}$, $v_{max}=0.768$, the grid density is $2^{15}$ points in each direction.}
\label{fig:v_vs_r}
\end{figure}

The first feature of the collapsing field  is the formation, or not, of a
black hole.  To examine this we plot a diagram of $v$ vs $r$, the area coordinate, for different 
values of $u$ (See figure \ref{fig:v_vs_r}). Each line represents an outgoing null ray of a constant $u$, namely a trajectory  of a photon emitted from the origin at  $u$. Rays with $u$ small enough show a flat-like spacetime behavior, for which $v \approx u+2r$.
For small values of $u$, at early times, the rays don't encounter a strong gravitational field and they escape 
to infinity.
At later times the gravitational field becomes stronger and the outgoing rays are bend more and more before 
they eventually manage to escape. Once a black hole forms these rays are trapped. For  subcritical initial configuration the field disperses and all outgoing null rays reach infinity (see figure \ref{fig:noblack hole}).
For a supercritical initial configuration a horizon appears when  an outgoing null ray doesn't  escape and doesn't reach future null infinity
but rather remains in the same radius $r$ for all values of $v$.
Later rays that emerge from the origin  collapse back to the origin (see figure \ref{fig:black hole}). 

\begin{figure}[!ht]
 \centering
\includegraphics[width=0.85\textwidth]{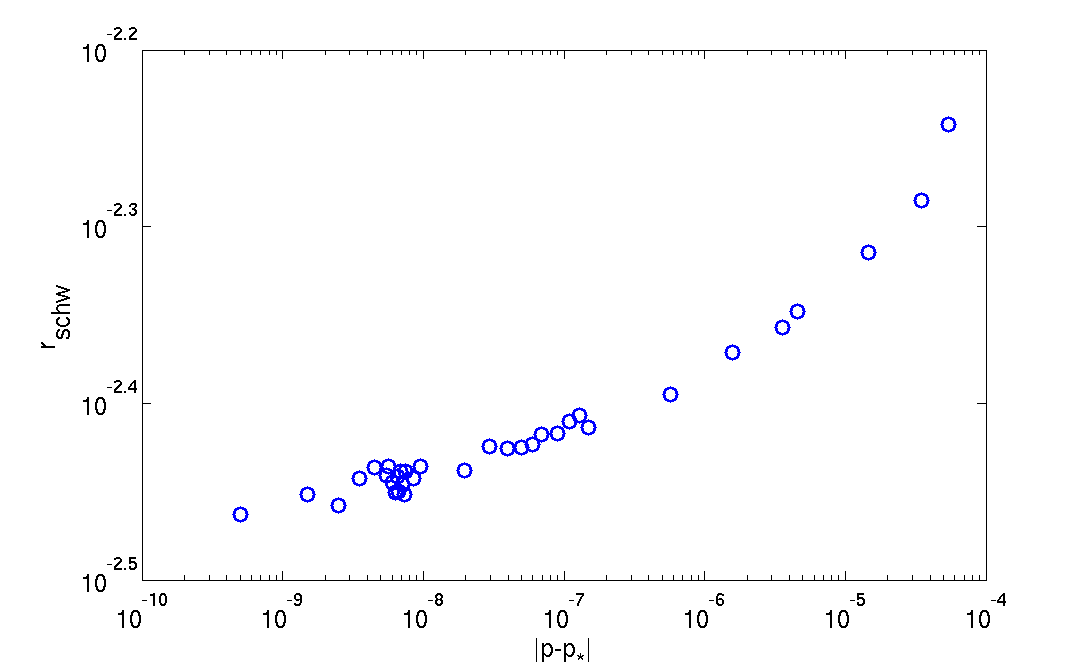}
\caption{\footnotesize 
The  black hole's Schwarzschild radius - $r_{s}$ vs.  the difference between 
the amplitude  and the critical one - $|p-p_*|$, 
for supercritical cases with $\alpha=10^{-4}$.
In the classical GR solution this relation is a power law. Here, in EGB  gravity,
a power law is not observed.
}
\label{fig:r0_vs_p-pStar}
\end{figure}

While, as expected the Gauss-Bonnet terms don't change the overall classical GR behavior, the critical behavior is lost. 
For classical GR in $D=5$ dimensions the black hole radius scales as $R \propto(p-p_*)^{\gamma}$.
Figure \ref{fig:r0_vs_p-pStar} depicts the black hole radius as a function of $|p-p_*|$ 
for supercritical evolutions with $\alpha=10^{-4}$. Note that we use here  the scaling relation 
of the Schwarzschild radius instead of the mass (see section \ref{sec:black hole_mass}).
The dependence of the black hole radius on the initial amplitude is monotonic, i.e. for larger values of $p$ the black hole
radius is larger. However in the EGB gravity we don't observe the classical power-law relation.
The existence of the  scaling relation is related to the  self similar properties of the classical solution \citep{Hod&Piran, Gundlach:gr-qc0210101}. Since self similarity is not preserved in EGB gravity we expect  that the  scaling relation will also be violated. 

\begin{figure}[t!]
 \centering
\includegraphics[width=0.8\textwidth]{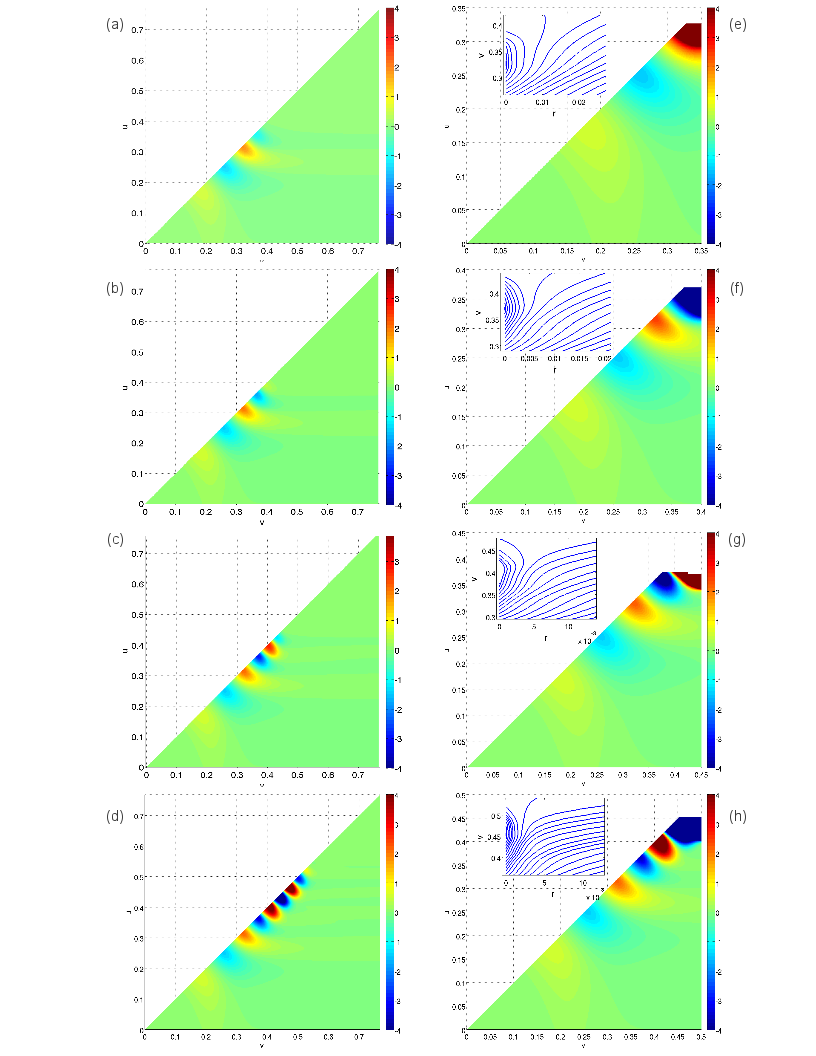}
\caption{\footnotesize  Contours of the scalar field function $s$ 
in EGB gravity with a coupling constant $\alpha=10^{-5}$, 
grid density $2^{15}$, $v_{max}=0.768$, $v_c=0.22$.
The initial field amplitude increases or decreases towards the critical amplitude: $p\rightarrow p_*$.
Plots (a-d) - subcritical collapse: $p < p_*$. 
Plots (e-h) - supercritical collapse $p > p_*$. The inserts  demonstrate 
the formation of black holes as seen in the $v-r$ diagram.
The initial values of the  field amplitude are: $p=0.1654$(a), $0.16557$(b), $0.1657$(c), $0.165723$(d),
 $0.1676$(e), $0.166$(f), $0.1658$(g), $0.16573$(h).}
\label{fig:app_critical}
\end{figure}

In classical GR the solution is discretely self similar (DSS) just below the black hole threshold. 
The field reaches the origin, oscillates and then disperses or collapses, depending on whether it is subcritical or supercritical. 
These oscillations don't depend on initial conditions and they decay in a DSS pattern (see figure \ref{fig:s_surf_0216}).
The addition of  the Gauss-Bonnet terms  destroys this behavior. 
With these terms the field still oscillates, but the self similarity disappears 
(See figure \ref{fig:app_critical}).
The oscillations grows at first  as if the additional terms in the Lagrangian amplify the field,  prevent it from decaying and keep the oscillations alive for a longer time. Eventually the oscillations decay and the field disperses.

In the classical GR solution,  we  observe more and more self similar transients when approaching the critical amplitude from both sides. 
In EGB solution,  while approaching the critical amplitude more and more oscillations are also 
observed. However, these oscillations are  not self similar and their scale does not decrease. 
Figure \ref{fig:app_critical} depicts  the contours of the scalar field $s$ for a set of subcritical solutions (panels (a-d)), with $\alpha=10^{-5}$, 
with growing amplitudes approaching the critical one,  and a set of supercritical solutions (panels (e-h)) with amplitudes decreasing
towards the critical one. 
An increasing number of oscillations is observed as $p$ approaches $p_*$.  
The  inserts   in the supercritical solution, 
figures \ref{fig:app_critical}(e-h), depict the $v-r$ diagram, demonstrating  black holes formation.
A cutoff in field diagram is a sign for a singularity 
(see the scheme in figure \ref{fig:penrose}).

\begin{figure}[!ht]
 \centering
\includegraphics[width=0.5\textwidth]{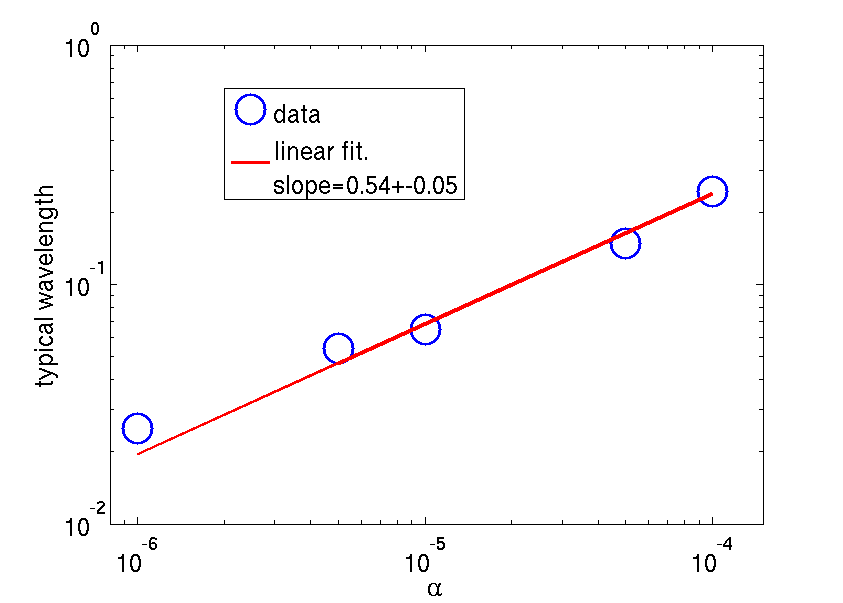}
\caption{\footnotesize 
The  typical wavelength of scalar field oscillations 
in EGB gravity near the black hole threshold vs. the coupling constant $\alpha$.
Red line indicates a linear fit with a slope: $m=0.54\pm0.05$.
Namely, the wavelength is proportional to $\sqrt{\alpha}$ as expected from the dimensional analysis.
}
\label{fig:lamda_alpha}
\end{figure}

\begin{figure}[b!]
 \centering
\includegraphics[width=0.7\textwidth]{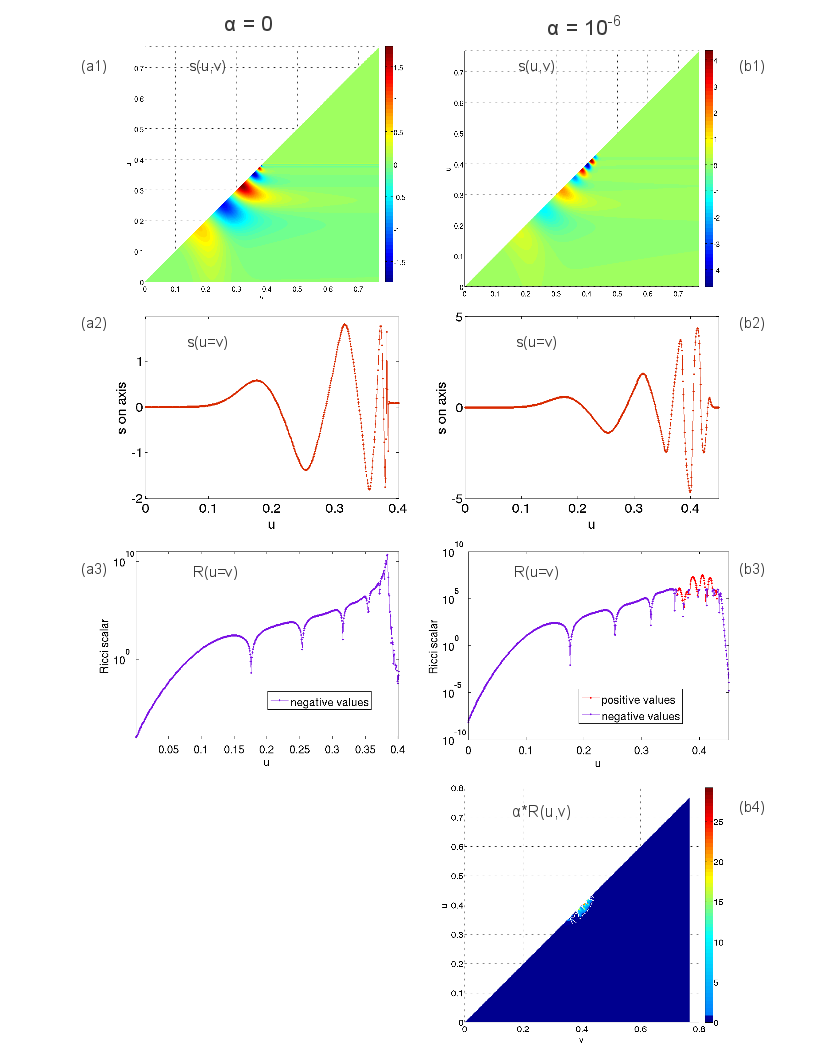}
\caption{\footnotesize 
A slightly subcritical collapse with different values of the coupling
constant $\alpha$.
Panels (a1-3) correspond to a classical GR solution, 
$\alpha=0$.
A self similar behavior can be observed. Panels (b1-4) correspond to EGB gravity  with 
$\alpha=10^{-6}$.
The first row, panels (a1) and (b1), presents contour plots of the  field function $s$. 
The second row, panels (a2) and (b2), shows the field function $s$ at the origin ($r=0$) vs. $u$.
The third row, panels (a3) and (b3), shows the Ricci scalar on the origin vs. $u$ in a logarithmic scale. 
A red color indicates positive values and a purple color negative values of $R$. 
Panel (b4), displays contour plot of $|\alpha R|$, showing
the regions where the  higher order terms are significant, i.e. $|\alpha R|>1$.
 }
\label{fig:alpha_0_minus6}
\end{figure}

\begin{figure}[t!]
 \centering
\includegraphics[width=0.7\textwidth]{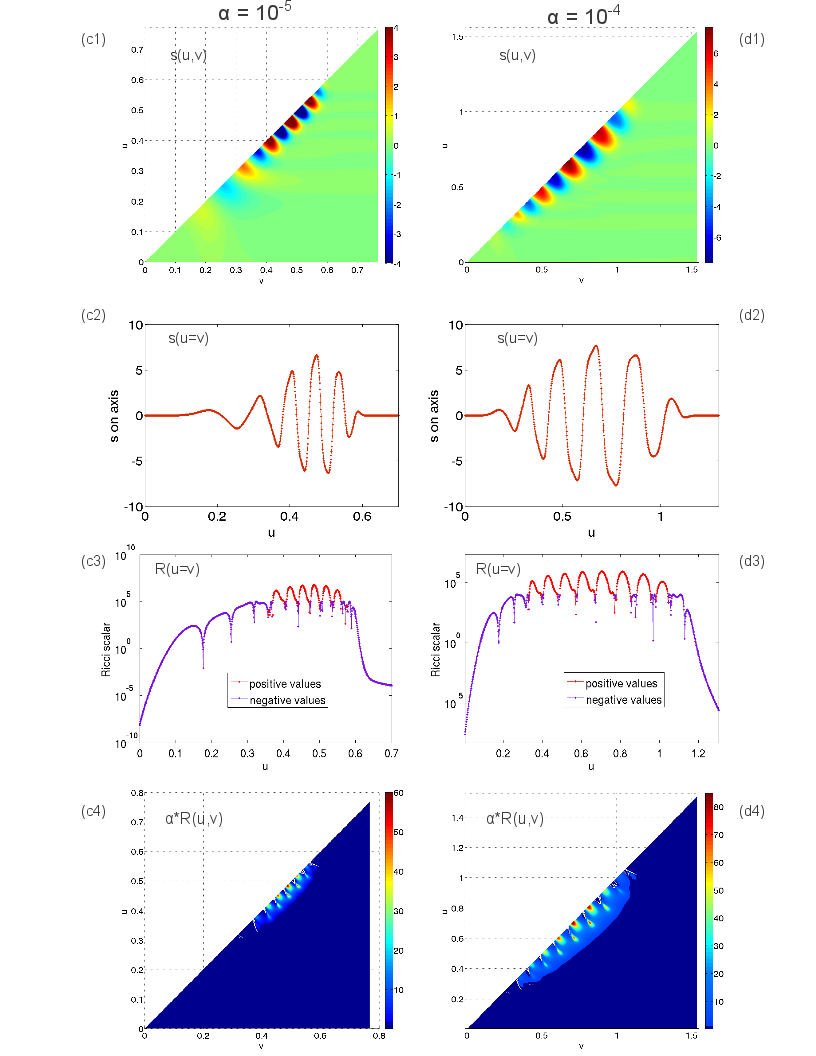}
\caption{\footnotesize A continuation of figure \ref{fig:alpha_0_minus6} 
with larger values of $\alpha$ in a slightly
subcritical collapse. 
Panels (c1-4) correspond to EGB gravity solution with 
$\alpha=10^{-5}$.
Panels (d1-4) correspond to EGB gravity solution with 
$\alpha=10^{-4}$.
}
\label{fig:alpha_minus5_minus4}
\end{figure}

Obviously the self similarity disappear not only in the field functions, but in all metric functions and 
their derivatives. 
Particularly interesting is the behavior of the Ricci scalar (equation (\ref{R_axes})) 
as seen in  Figure \ref{fig:alpha_0_minus6} panels (a3) and (b3), and
Figure \ref{fig:alpha_minus5_minus4} panels (c3) and (d3).
This figures show the Ricci scalar at the origin, $R(r=0)$,  as a function of $u$  for  different values of  $\alpha$. 
Purple color indicates negative values and red indicates positive  values of the Ricci scalar. 
Figure \ref{fig:alpha_0_minus6}(a3) shows the classical solution, i.e. $\alpha=0$. In classical GR
the condition $\alpha<\frac{45a^2}{16z^2}$ (equation \ref{eq:ricciCondition}) is always satisfied, 
thus $R(r=0)$ is always negative and it never changes sign.
On the other hand for EGB gravity this condition is inevitably violated 
for amplitudes close enough to the critical one (see figures \ref{fig:alpha_0_minus6}(b3), 
\ref{fig:alpha_minus5_minus4}(c3) and \ref{fig:alpha_minus5_minus4}(d3)).
As $p$ approaches $p_*$, the metric function $a$  tends to zero and the derivative of the field  $z$ grows. 
At some point the condition (\ref{eq:ricciCondition}) is violated and the Ricci scalar changes sign, indicating a change 
in the local geometry. This demonstrates that near the black hole threshold  for $\alpha>0$ the local geometry is different from 
the classical GR geometry.

Figures \ref{fig:alpha_0_minus6} and \ref{fig:alpha_minus5_minus4} 
compare slightly subcritical solutions 
with different values of the coupling constant, $\alpha=0,10^{-6},10^{-5},10^{-4}$ 
in each column (columns a,b,c and d).
The first, upper row of each figure (panels a1, b1, c1 and d1) shows contours of the field function $s$.
The second and the third rows (panels a2-3, b2-3, c2-3 and d2-3) show the field function $s$ and the 
Ricci scalar on the origin ($r=0$) vs. $u$.
The $4^{th}$ row (panels b4, c4 and d4) shows a contour plot of $|\alpha R|$.
It indicates the regions where the additional curvature terms are significant, i.e. 
regions where $|\alpha R|\geq1$. Naturally with larger values of  $\alpha$  this region grows and $|\alpha R|$ reaches larger values.  

The plots of the field function - $s$ - on the origin (b2, c2 and d2) 
nicely show a ``beat``-like pattern, increasing and then decreasing, in the field pulsation's strength. 
However, for small values of $\alpha$, (see figure \ref{fig:alpha_0_minus6}(b2)), 
the field behavior on the origin resembles, at least
initially, a self similar behavior in regular GR. 
This could be explained by the low and insignificant values of the higher order  terms 
in these regions (see figure \ref{fig:alpha_0_minus6}(b4)). At the same time a comparison  of the field $s$ and the 
Ricci scalar at the origin for different values of $\alpha$
(panels (a2-3), (b2-3), (c2-3) and (d2-3))
reveals  that the new ''beat`` form of the field oscillations appears at the same retarded time $u$ at 
which Ricci curvature changes sign and becomes positive.
%

The scalar field oscillations  (figures \ref{fig:app_critical}, \ref{fig:alpha_0_minus6} 
and \ref{fig:alpha_minus5_minus4}) show a typical wavelength which depends on the value of
 $\alpha$ (see  figure \ref{fig:lamda_alpha}). 
All the lengths are measured in the simulation length units $[u]$.
As expected from a dimensional  analysis the typical wavelength of the oscillations 
is proportional to $\sqrt{\alpha}$.

\section{Summary}
We have developed a numerical scheme for simulating  the dynamical collapse of a spherically symmetric massless 
scalar field  in EGB gravity. This model for gravity  includes higher (quadratic)  
order curvature corrections to the Hilbert action. These corrections induce changes 
in Einstein equations, which govern the evolution of the system when the curvature is  large.

We find  that the addition of higher order curvature correction 
destroys the classical Choptuik phenomenon.
The introduction of the dimensional coupling constant  $\alpha$, which has a units of $length^2$,
destroys the scale invariance of the system.  As a consequence
the self similar behavior, which is an integral part of the critical phenomena in regular GR, disappears. 
Instead the solution shows a different pattern of pulsations with a typical wavelength,
which is proportional to $\sqrt{\alpha}$, as expected from a dimensional analysis. 
The changes in the oscillations pattern are accompanied by changes in the sign of the Ricci scalar  at the origin, 
indicating a change in the local geometry of the space-time. 

We thank Shahar Hod and Nathalie Deruelle for many helpful discussions, Stanley Deser and Hideko Maeda for useful remarks and Yonatan Oren and Evgeny Sorkin for assistance with the numerical calculations.

\rule[-1cm]{5cm}{0.01cm}
%

\end{document}